\documentclass[onecolumn,amsmath,nofootinbib,secnumarabic,eqsecnum,amssymb,12pt]{revtex4}

\usepackage{bm}

\usepackage{epsf}
\usepackage{graphicx,epsfig}
\usepackage{amsfonts}
\usepackage{amssymb}
\usepackage{xcolor}
\usepackage{subfigure}

\definecolor{mine}{rgb}{0.2,0.1,0.7}
\definecolor{bb}{rgb}{0.3, 0.5, 1}
\definecolor{bg}{rgb}{0.1, 0.1, 0.5}
\usepackage[%
    pdftitle={The title of your letter},
    pdfauthor={Your name},
    pdfsubject={The subject of the letter},
    pdfkeywords={Any keywords},
     colorlinks=true,  
    linkcolor=mine,
     citecolor=bg, 
     urlcolor=bg, 
     pdfnewwindow=true,
     pagebackref=true,
     filecolor=bb,
     pdffitwindow=false,     
     pdfstartview={FitH}    
]{hyperref}

\def\M{M_P}
\def\z{\zeta}
\newcommand{\dn}[2]{{\mathrm{d}^{{#1}}{{#2}}}}

\newcommand{\bea}{\begin{eqnarray}}
\newcommand{\eea}{\end{eqnarray}}
\newcommand\be{\begin{equation}}
\newcommand\ee{\end{equation}}
\newcommand\beq{\begin{equation}}
\newcommand\eeq{\end{equation}}

\newcommand{\nn}{\nonumber}

\newcommand{\refeq}[1]{(\ref{#1})}
\def\Tdot#1{{{#1}^{\hbox{.}}}}

\def\p{\partial}
\def\epss{\epsilon_s}
\def\c{c_s}
\def\eps{\epsilon}
\def\etas{\eta_s}
\def\etaf{\eta_f}
\def\s{s}

\makeatletter
\renewcommand\section{\@startsection {section}{1}{\z@}%
                                 {-3.5ex \@plus -1ex \@minus -.2ex}
                                   {2.3ex \@plus.2ex}%
                                   {\normalfont\large\bfseries}}
\renewcommand\subsection{\@startsection{subsection}{2}{\z@}%
                                   {-3.25ex\@plus -1ex \@minus -.2ex}%
                                     {1.5ex \@plus .2ex}%
                                     {\normalfont\bfseries}}
\renewcommand\subsubsection{\@startsection{subsubsection}{3}{\z@}%
                                   {-3.25ex\@plus -1ex \@minus -.2ex}%
                                     {1.5ex \@plus .2ex}%
                                     {\normalfont\bf\itshape}}
                                    
\makeatother

\begin{document}

\begin{center}
{\Large \bf 
On the redundancy of operators and the bispectrum \\
\vspace{0.2cm}
in the most general second-order scalar-tensor theory \\
}

\vskip 1.3cm
\centerline{\large S\'ebastien Renaux-Petel 
}

\vskip 0.8cm
{\em    Centre for Theoretical Cosmology,\\
Department of Applied Mathematics and Theoretical Physics,\\
University of Cambridge, Cambridge CB3 0WA, UK    \\[0.3in]}

\end{center}

\begin{center}
{\bf
Abstract}
\end{center}

In this short note we explain how to use the linear equations of motion to simplify the third-order action for the cosmological fluctuations. No field redefinition is needed in this exact procedure which considerably limits the range of independent cubic operators, and hence of possible shapes of the primordial bispectrum. We demonstrate this in the context of the most general single-field scalar-tensor theory with second-order equations of motion, whose third-order action has been calculated recently in arXiv:1107.2642 and 1107.3917. In particular, we show that the three cubic operators initially pointed out in these works as new compared to $k$-inflation can actually be expressed in terms of standard $k$-inflationary operators.

\noindent

\setcounter{page}{1}

\section{I\lowercase{ntroduction}}

The deviation from perfect Gaussian statistics of the primordial fluctuations promises to allow a more precise discrimination between competing scenarios of the early Universe than has hitherto been possible. A central object in this respect is the primordial bispectrum: the three-point correlation function of the primordial curvature perturbation $\z$ in Fourier space $\langle \z (\boldsymbol{k}_1)  \z (\boldsymbol{k}_2)  \z (\boldsymbol{k}_3)\rangle$. An easy way to acknowledge the wealth of information contained in this object is to note that it is a function of three variables, contrary to the power spectrum which is a function of only one variable. Consequently, considerable amount of attention has been given to understand and classify the bispectrum's dependence on the configuration of the momenta $\boldsymbol{k}_1,\boldsymbol{k}_2,\boldsymbol{k}_3$, what is called its shape \cite{Babich:2004gb,Fergusson:2008ra}. Beyond the standard local \cite{Komatsu:2001rj} and equilateral \cite{Creminelli:2005hu} shapes of non-Gaussianities, lots of other patterns, corresponding to different mechanisms for generating the primordial fluctuations, have now been identified (see for instance \cite{Chen:2010xk} and \cite{Liguori:2010hx} for theoretical and observational reviews respectively). In practice, shapes are derived from operators in the third-order action for the fluctuations, so that a natural question to ask is: what is the set of such possible cubic operators?  In this respect, we elaborate in this short note on a technical point whose important consequences have not yet been fully acknowledged: some cubic operators that are \textit{a priori} independent, \textit{i.e.} not simply related by integrations by part, can actually be related to each other by using the linear equations of motion, a procedure which has recently been stressed to be valid for computing correlation functions \cite{Arroja:2011yj,Burrage:2011hd}. This considerably limits the range of truly independent cubic operators, and hence of possible shapes of non-Gaussianities.

We demonstrate this in the framework of Hordenski's most general single-field scalar-tensor theory with second-order equations of motion \cite{Hordenski}, which has been ``resurrected'' recently  \cite{Charmousis:2011bf} and which has been shown to be equivalent to generalized Galileons in four dimensions \cite{Deffayet:2011gz} in reference \cite{Kobayashi:2011nu}. The corresponding third-order action has been calculated in references \cite{Gao:2011qe,DeFelice:2011uc}, where the appearance of three cubic operators absent in the simpler class of models known as $k$-inflation was stressed. As we pointed out to Gao and Steer for a revised version of their preprint \cite{Gao:2011qe}, and which we now explain in further details, these ``new'' operators can actually be expressed in terms of ones which have already been studied in depth in the context of $k$-inflation \cite{Burrage:2011hd}, thereby alleviating the need for their separate study. 

In the next section, we quote the second-and third-order action for scalar perturbations in the most general single-field scalar-tensor theory with second-order equations of motion and we explain the strategy behind our explicit calculations in section \ref{Calculations}. We 
briefly conclude in section \ref{Conclusion}.

\section{T\lowercase{he most general second-order scalar-tensor theory}}

The action of the most general four-dimensional theory involving a metric $g_{\mu \nu}$ and a scalar field $\phi$ and whose equations of motion are of second order can be written as \cite{Hordenski,Deffayet:2011gz,Kobayashi:2011nu} (we use units in which $\hbar=c=\M=1$)
\begin{equation}{\label{action}}
        S = \int d^4x \sqrt{-g} \left( \frac{1}{2}R + \sum_{n=0}^{3}\mathcal{L}_{n} \right)
    \end{equation}
where
\begin{eqnarray}
      \mathcal{L}_{0} & = & K^{}\left(X,\phi\right),\\
        \mathcal{L}_{1} & = & G^{(1)}\left(X,\phi\right)\Box\phi,\\
        \mathcal{L}_{2} & = & G_{,X}^{(2)}\left(X,\phi\right)\left[\left(\Box\phi\right)^{2}-\left(\nabla_{\mu}\nabla_{\nu}\phi\right)^{2}\right]
        +R\, G^{(2)}\left(X,\phi\right) ,\\
        \mathcal{L}_{3} & = & G_{,X}^{(3)}\left(X,\phi\right)\left[\left(\Box\phi\right)^{3}-3\Box\phi\left(\nabla_{\mu}\nabla_{\nu}\phi\right)^{2}
        +2\left(\nabla_{\mu}\nabla_{\nu}\phi\right)^{3}\right]-6G_{\mu\nu}\nabla^{\mu}\nabla^{\nu}\phi\,
        G^{(3)}\left(X,\phi\right) .
        \end{eqnarray}
Here, $K(X,\phi)$ and $G^{(n)}(X,\phi)$ are arbitrary functions of $\phi$ and of the kinetic term 
\be
X \equiv -\frac12 g^{\mu \nu} \p_{\mu} \phi \p_{\nu} \phi\,,
\ee
the case of $k$-inflation \cite{ArmendarizPicon:1999rj,Garriga:1999vw} being recovered when $G^{(1,2,3)}=0$\footnote{Accurately, $k$-inflation is also recovered when $G^{(2,3)}=0$ and $G^{(1)}$ is a function of $\phi$ only.}; $R$ and $G_{\mu \nu}$ denote respectively the Ricci scalar and Einstein tensor associated to the metric $g_{\mu \nu}$ and $\Box \phi \equiv \nabla^{\mu} \nabla_{\mu} \phi$, $\left(\nabla_{\mu}\nabla_{\nu}\phi\right)^{2} \equiv \left(\nabla_{\mu}\nabla_{\nu}\phi\right)\left(\nabla^{\mu}\nabla^{\nu}\phi\right)$ and $\left(\nabla_{\mu}\nabla_{\nu}\phi\right)^{3} \equiv  \left(\nabla_{\mu}\nabla_{\nu}\phi\right)\left(\nabla^{\mu}\nabla^{\rho}\phi\right)\left(\nabla_{\rho}\nabla^{\nu}\phi\right)$\,. One should point out that only some particular models of this type are stable under radiative corrections (see for example \cite{Hinterbichler:2010xn}). However, we will consider this general class of models in the following, as it is of no more computational cost, and actually simpler, than considering algebraically special cases.

The interested reader can find the explicit form of the corresponding equations of motion in references \cite{Gao:2011mz,Kobayashi:2011nu}. Our purpose here is to consider cosmological perturbations about a spatially flat Friedmann-Lema\^itre-Robertson-Walker spacetime of metric
\be
ds^2=-dt^2+a^2(t) d{\boldsymbol{x}}^2\,,
\ee
where $t$ is cosmic time. The only scalar degree of freedom can be conveniently chosen to be the curvature perturbation $\z$, defined such that the spatial metric takes the form
\be
g_{ij}=a^2 e^{2 \zeta} \delta_{ij}
\ee
in the uniform inflaton gauge in which $\delta \phi=0$ (we do not consider gravitational waves here). The second-order action in terms of $\zeta$ then takes the form
   \begin{equation}{\label{S2}}
        S_{(2)} = \int {\rm d}t \,\dn{3}{x}\, a^3  \frac{\epss}{\c^2}  \left({\dot \z}^2 - \c^2 \frac{(\partial \zeta)^2}{a^2}\right)
    \end{equation}
    where the explicit expressions of $\epss$ and $\c^2$, that will be unimportant here, can be found in \cite{Gao:2011qe,Kobayashi:2011nu}. From this, one deduces the linear equation of motion for $\z$:
\be
\frac{1}{a^3 \epss} \Tdot{\left( \frac{a^3 \epss}{\c^2} \dot \zeta \right)}=\frac{\partial^2 \zeta}{a^2}\,.
\label{mode-equation}
\ee
Finally, the third-order scalar action given in reference \cite{Gao:2011qe} (see also \cite{DeFelice:2011uc}) reads
   \begin{eqnarray}
        S_{(3)} & = &    \int {\rm d}t \,\dn{3}{x}\, a^3   \biggl\{\frac{\Lambda_{1}}{H} \dot \zeta ^{3}+\Lambda_{2} \z \dot \zeta^{2}+\Lambda_{3}\zeta\frac{(\p \z)^2}{a^2}+\frac{\Lambda_{4}}{H^{2}} \dot \zeta^{2} \frac{ \partial^{2}\zeta}{a^2}+ \Lambda_{5} \dot \zeta \partial_i \zeta \partial^i \left( \p^{-2} \dot \z \right) + \Lambda_{6}\partial^{2}\zeta  \left(\partial_i \p^{-2} \dot \z  \right)^{2}\nonumber \\
         &  & +\frac{\Lambda_{7}}{a^4 H^{2}}\left[   \left(\partial \zeta\right)^{2} \p^2 \z-\zeta\partial_{i}\partial_{j}\left(\partial^{i}\zeta\partial^{j}\zeta\right)\right] + \frac{\Lambda_{8}}{a^2 H}\left[   \p^2 \z \p_i \z \p^i \p^{-2} \dot \z  -\z \p_i \p_j (\p^i \z \p^j  \p^{-2} \dot \z)      \right]
         \biggr\},\label{S3}
        \end{eqnarray}       
where $H \equiv \frac{\dot a}{a}$ is the Hubble parameter, the $\Lambda_i$'s are dimensionless time-dependent parameters and we discarded the term proportional to the linear equation of motion \refeq{mode-equation} that was given in references \cite{Gao:2011qe,DeFelice:2011uc} in addition to Eq. \refeq{S3}. Indeed, for the purpose of computing correlation functions of $\zeta$, this kind of terms are irrelevant, as was explained in \cite{Seery:2005gb,Seery:2006tq,Seery:2010kh} and emphasised recently in \cite{Arroja:2011yj,Burrage:2011hd}: they do not contribute to any Feynman graph at any order in perturbation theory because $\delta S_{(2)} / \delta \z$ is zero by construction when evaluated on a propagator. A straightforward consequence is that we are free to simplify the third-order action \refeq{S3} by using the linear equation of motion \refeq{mode-equation}. This is what we will use in the next section to show that the operators in $\Lambda_4, \Lambda_7$ and $\Lambda_8$ are redundant\footnote{See section 7.7 of \cite{Weinberg:1996kr} for a related discussion about redundant couplings in quantum field theory in Minkowski spacetime.}. Before that, we should point out two subtleties: first, some temporal boundary terms, generated for instance by integrations by part, can give a non-zero contribution to late-time correlation functions of $\zeta$ \cite{Seery:2006tq,Arroja:2011yj,Burrage:2011hd}. No such dangerous terms will arise in the calculations below, in which temporal boundary terms are therefore dismissed\footnote{Cosmological observables are only related to late-time correlation functions of $\zeta$. However, should one want to calculate the bispectrum at any time, these boundary terms would have to be taken into account.}. However, we should bear in mind that it may not be the case of the boundary terms that were not kept by the authors of references \cite{Gao:2011qe,DeFelice:2011uc} in the calculations that led to Eq. \refeq{S3}. Second, even for the comparatively simpler case of $k$-inflation, different forms of the third-order action are given in different references, precisely because of the freedom to use the linear equation of motion to simplify it. For example, the well known form given in references \cite{Seery:2005wm,Chen:2006nt} contains some operators absent in Eq. \refeq{S3}. However, boundary terms must be taken into account in that case. In the following, when we refer to the third-order action of $k$-inflation, we have in mind the one given recently in reference \cite{Burrage:2011hd} (in Eq. 3.10). It is particularly convenient because it is of the form \refeq{S3} -- with $\Lambda_{4,7,8}=0$ -- and that no boundary terms are then necessary.

\section{R\lowercase{eplacing redundant operators}} 
\label{Calculations}

In this section, we show in some details how to use the linear equation of motion \refeq{mode-equation} to replace the cubic operators in $\Lambda_4, \Lambda_7$ and $\Lambda_8$ by the others in Eq. \refeq{S3}. We will use the parameters
\be
\epsilon \equiv -\frac{\dot H}{H^2}\,,\quad \etas \equiv \frac{\dot \epss}{H \epss}\,, \quad s \equiv \frac{\dot \c}{H \c}\,, \quad \etaf \equiv \frac{\dot f}{H f}
\label{SV1}
\ee
\be
\eta_4 \equiv \frac{\Tdot{\left(\frac{\Lambda_4}{H^2} \right)}}{H \left(\frac{\Lambda_4}{H^2} \right)}\,, \quad
\eta_7 \equiv \frac{\Tdot{\left(\frac{\Lambda_7}{H^2} \right)}}{H \left(\frac{\Lambda_7}{H^2} \right)}\,, \quad 
\eta_8 \equiv \Tdot{\left(\frac{\Lambda_8}{H} \right)}/ \Lambda_8
\label{SV2}
\ee
which are neither considered to be small nor slowly varying, \textit{i.e.} they are only short-hand notations. Here and in the following, $f(t)$ stands for any time-dependent function.

\begin{itemize}
\item  {\bf Operator $\dot \z^2 \p^2 \zeta$ ($\Lambda_4$):}
\end{itemize}

Using the linear equation of motion \refeq{mode-equation} to replace $\partial^2 \zeta$ and integrating by parts twice, one readily obtains
\beq
\fbox{$\displaystyle
 \int {\rm d}t \,\dn{3}{x}\, a  f    \dot \zeta^2  \partial^2 \zeta=\int {\rm d}t \,\dn{3}{x}\, a^3 \frac{2 H f}{\c^2} \dot \zeta^3 \left(1+ \frac{\etas}{2}-\frac{2\s}{3}-\frac{\etaf}{6}    \right)\,. $}
 \label{1}
\eeq
Hence, the operator $ \dot \zeta^2  \partial^2 \zeta$ can be replaced by the operator $\dot \z^3$.

\begin{itemize}
\item  {\bf Operators $(\p \z)^2 \p^2 \zeta$ and $\zeta\partial_{i}\partial_{j}\left(\partial^{i}\zeta\partial^{j}\zeta \right)$ ($\Lambda_7$):}
\end{itemize}

First, let us note that simple spatial integrations by part gives $\int \zeta\partial_{i}\partial_{j}\left(\partial^{i}\zeta\partial^{j}\zeta \right)=-\frac12 \int (\p \z)^2 \p^2 \zeta$. Hence, we only have to deal with the latter operator. Using the same steps as above as well as the spatial integration by parts $\int \dot \zeta \p \zeta \p \dot \zeta =-\int \frac12  \dot \zeta^2 \partial^2 \zeta$, one gets\footnote{We discarded a boundary term here in $a \dot \zeta (\p \zeta)^2$. Although it is linear in $\dot \zeta$, and hence potentially dangerous according to the criterion given in \cite{Burrage:2011hd}, one can check that it does not contribute the primordial bispectrum.}
\bea
\int {\rm d}t \,\dn{3}{x}\, \frac{f}{a}  (\p \zeta)^2  \partial^2 \zeta&=&\int {\rm d}t \,\dn{3}{x}\, a \frac{f}{\c^2} \left[ \dot \zeta^2 \partial^2 \zeta+ H \dot \zeta (\p \zeta)^2 \left(2+\etas-\etaf \right) \right]  \,,
\label{2'}
\eea
which, together with Eq. \refeq{1}, gives
\beq
\fbox{$\displaystyle
\int {\rm d}t \,\dn{3}{x}\, \frac{f}{a}  (\p \zeta)^2  \partial^2 \zeta=\int {\rm d}t \,\dn{3}{x}\, a^3 \frac{f H}{\c^4} \left[ \dot \zeta^3 \left(1+\etas-\frac{2\s}{3} -\frac{\etaf}{2}  \right)+ \c^2  \dot \zeta  \frac{(\p \zeta)^2}{a^2} \left(2+\etas-\etaf \right)  \right]  $}\, .
\label{2}
\eeq
\begin{itemize}
\item  {\bf Operator $\dot \z (\p \z)^2$:}
\end{itemize}

At leading order in a slow-varying approximation, in which all the parameters in Eqs. \refeq{SV1}-\refeq{SV2} can be neglected, results similar to \refeq{1}-\refeq{2} were used recently in references \cite{Mizuno:2010ag,Creminelli:2010qf,Burrage:2010cu,RenauxPetel:2011dv,RenauxPetel:2011uk}. However, Eq. \refeq{2} is not completely satisfactory for our purpose here because the operator $\dot \z (\p \z)^2$ in its right-hand side does not appear in the form \refeq{S3} of the third-order action. Fortunately, we now show that $\dot \z (\p \z)^2$ itself can be expressed in terms of standard $k$-inflationary operators: spatially integrating by parts the term in $(\p \zeta)^2$ and subsequently integrating by parts in time give
\beq
\int {\rm d}t \,\dn{3}{x}\, a f \dot \zeta  (\p \zeta)^2 =\int {\rm d}t \,\dn{3}{x}\, a f \left(  H \zeta (\p \zeta)^2 \left(1+\etaf \right) -2 \zeta \dot \zeta \p^2 \zeta \right) \,.
\label{3'}
\eeq
Replacing $\p^2 \zeta$ in the last term of Eq. \refeq{3'} by using the linear equation of motion \refeq{mode-equation} and further integrating by parts in time, one then finds
\beq
\fbox{$\displaystyle
\int {\rm d}t \,\dn{3}{x}\, a f \dot \zeta  (\p \zeta)^2 =\int {\rm d}t \,\dn{3}{x}\, a^3 \frac{f H}{\c^2} \left(\frac{\dot \zeta^3}{H}- \z \dot \z^2 \left(3+2\etas-2\s-\etaf \right)+   \c^2 \zeta \frac{(\p \zeta)^2}{a^2} \left(1+\etaf \right)
 \right)$}
 \label{3}
\eeq
where the three operators on the right-hand side are the first three ones in Eq. \refeq{S3}.\\

\begin{itemize}
\item  {\bf Operators $\p^2 \z \p_i \z \p^i \p^{-2} \dot \z$ and $\z \p_i \p_j (\p^i \z \p^j  \p^{-2} \dot \z) $ ($\Lambda_8$):}
\end{itemize}

First, similarly to the second operator in $\Lambda_7$, simple spatial integrations by part gives $\z \p_i \p_j (\p^i \z \p^j  \p^{-2} \dot \z)=-\frac12 \int \dot \z (\p \z)^2$, which we know how to treat with Eq. \refeq{3}. For the operator $\p^2 \z \p_i \z \p^i \p^{-2} \dot \z$: replacing $\p^2 \zeta$ by using the linear equation of motion \refeq{mode-equation}, integrating by parts in time, using the linear equation of motion again to express $\ddot \z$ in terms of $\dot \z$ and $\p^2 \z$ and the spatial integration by part $\int \dot \z \p_i \dot \z \p^i \p^{-2} \dot \z=-\int \frac12 \dot \z^3$, one finds\\

\fbox{
 \begin{minipage}{\linewidth}
\bea
\int {\rm d}t \,\dn{3}{x}\, a f \p^2 \z \p_i \z \p^i \p^{-2} \dot \z  &=&\int {\rm d}t \,\dn{3}{x}\, a^3 \frac{f}{\c^2} \left[  \frac{\dot \z^3}{2}-\c^2\dot \z \frac{(\p \z)^2}{a^2}
\right.
\cr
&&
\left.
\hspace*{0.0em}+H \dot \z \p_i \z \p^i \p^{-2} \dot \z   \left(3+2\etas-2s-\etaf \right)  
 \right] 
\eea
\end{minipage}
}
\\

\noindent where again the operator $\dot \zeta  (\p \zeta)^2$ can be further expressed in terms of $\dot \z^3$, $\z \dot \z^2$ and $\z (\p \z)^2$ using Eq. \refeq{3}.

\begin{itemize}
\item  {\bf Simplifying the third-order action:}
\end{itemize}

Using the results above, we have all the ingredients to rewrite the third-order action \refeq{S3} in terms of only the well known operators that appear in the form of the $k$-inflationary third-order action given in reference \cite{Burrage:2011hd}. Effectively, this means that the terms in $\Lambda_{4,7,8}$ are redundant and that we can use the same form of third-order action as Eq. \refeq{S3} with the following redefined coefficients:

\bea
\tilde \Lambda_1&=&\Lambda_1+\frac{ \Lambda_4}{3\c^2} \left(6+ 3 \etas -4\s-\eta_4   \right)+\frac{\Lambda_7}{\c^4}\left(6+3\etas-\s-2\eta_7 \right) \\
\tilde \Lambda_2&=&\Lambda_2  -\frac{3\Lambda_7}{2\c^4} \left( (3+\eps+2\etas-\eta_7)(2+\etas-\eta_7)-\frac{\dot \etas}{H}+\frac{\dot \eta_7}{H} \right)  \nn 
\cr
&&
\hspace*{+1.5em}+\frac{\Lambda_8}{2 \c^2}\left(3+2\etas-2\s-\eta_8 \right) \\
\tilde \Lambda_3&=&\Lambda_3   + \frac{3\Lambda_7}{2\c^2} \left( (1-\eps-2\s+\eta_7)(2+\epss-\eta_7)+\frac{\dot \etas}{H}-\frac{\dot \eta_7}{H} \right)-\frac{\Lambda_8}{2}(1+\eta_8) \\
\tilde \Lambda_5&=&\Lambda_5+\frac{\Lambda_8}{\c^2} \left(3+2\etas-2\s-\eta_8 \right)\\
\tilde \Lambda_6&=&\Lambda_6\,, \quad \tilde \Lambda_4=0\,, \quad \tilde \Lambda_7=0\,, \quad \tilde \Lambda_8=0\,.
\eea

Of course, an immediate application is that the shapes of the bispectrum associated to the operators in $\Lambda_{4,7,8}$ that were calculated in references \cite{Gao:2011qe,DeFelice:2011uc} are linear combinations of the shapes associated to the operators in $\Lambda_{1,2,3,5}$, as it was already acknowledged for the term in $\Lambda_4$, and which we explained to Gao and Steer for a revised version of their preprint \cite{Gao:2011qe}. However, the corresponding correlation functions were then calculated at leading order in a slow-varying approximation, whereas we stress that our result is exact and is valid directly at the level of the action.

\section{C\lowercase{onclusions}}
\label{Conclusion}

In this short note, we have reconsidered the status of the operators that appeared in references \cite{Gao:2011qe,DeFelice:2011uc} in the third-order scalar action of the most general scalar-tensor theory with second-order equations of motion. There, three operators were identified besides the five ones that naturally appear in the simpler class of models known as $k$-inflation and whose properties are well known. Emphasizing that it is valid for computing correlation functions to use the linear equation of motion to simplify the third-order action, we have shown that these three operators are actually redundant and can be exactly traded for the five well known ones.

Under the slow-varying approximation and in the Bunch-Davies vacuum, these five independent cubic operators generate primordial bispectrum shapes either of equilateral type -- for $\dot \z^3$, $\dot \zeta \partial_i \zeta \partial^i \left( \p^{-2} \dot \z \right)$ and $\partial^{2}\zeta  \left(\partial_i \p^{-2} \dot \z  \right)^{2}$ -- or of local type -- for $\z \dot \z^2$ and $\z (\p \z)^2$. However, even in these circumstances, we should point out that this does not entail that nothing else can be expected from single field models of inflation. For instance, the orthogonal shape of non-Gaussianities \cite{Senatore:2009gt}, which has never been shown to be generated in a particular $k$-inflationary model, was recently shown to emerge in a simple model of DBI Galileon inflation \cite{RenauxPetel:2011dv,RenauxPetel:2011uk}. The set of possible shapes of course gets even larger when features or/and non-Bunch Davies vacuum are considered (see for example \cite{Chen:2010bk} for a recent illustration).

Eventually, let us note that Eq. \refeq{3} expresses the fact that the operator $\dot \z (\p \z)^2$ -- which typically generates a primordial bispectrum predominantly correlated with the equilateral template -- can be replaced by a combination of the three operators $\dot \z^3$, $\z \dot \z^2$ and $\z (\p \z)^2$, the last two generating a bispectrum whose shape is predominantly correlated with the local template. This -- and more generally the fact that cubic operators that are \textit{a priori} independent can actually be related on using the linear equations of motion -- indicates that one should be cautious with inferring results for the global shape of the primordial bispectrum from the type of operators present in a given form of the third-order action. \\

\noindent  {\bf  Acknowledgments:} I would like to thank David Seery for useful comments on a draft version of this paper. I am supported by the STFC grant ST/F002998/1 and the Centre for Theoretical Cosmology.

\bibliography{Biblio}        

\end{document}